# On the fundamental definition of critical current in superconductors

E.F. Talantsev

Robinson Research Institute, Victoria University of Wellington, 69 Gracefield Road, Lower Hutt, 5040, New Zealand

E-mail: evgeny.talantsev@vuw.ac.nz

*Abstract*

Transport critical current, $I_c$, is usually defined in terms of a threshold electric field criterion, $E_c$, with the convention $E_c = 1$ µV/cm, chosen somewhat arbitrarily to provide "reasonably small" electric power dissipation in practical devices. Thus $I_c$ is not fundamentally determined. However, recently it was shown, that the *self-field* critical current of thin-film superconductors is indeed a fundamental property governed only by the London penetration depth. Here we reconsider the definition of critical current and resolve the apparent contradiction. We measure the field distribution across the width of 2G high-$T_c$ superconducting tapes as the transport current is increased to $I_c$. We identify a threshold current, $I_{c\_surfB}$, at which two physical events occur simultaneously: (i) an abrupt crossover from non-linear to linear dependence of the local surface magnetic flux density, $B_{surf}$, as a function of transport current measured at any point on the superconductor surface. This effect was not reported previously. (ii) the appearance of a non-zero electric field, just above of the sensitivity of measuring system. In the present examples $I_{c\_surfB}$ is 10-15% lower than $I_{c\_E}$ determined by the $E_c$ criterion. We propose the transition of $B_{surf}(I)$ from non-linear to linear as the most reliable and more fundamental technique for measuring transport critical currents.

# On the fundamental definition of critical current in superconductors

The transport critical current, $I_c$, is commonly defined as the current at which the induced electric field, $E$, reaches a certain critical value, $E_c$ [1]. The commonly accepted criterion is for $E_c = 1$ µV/cm [2,3]. There were several suggestions to define the critical current based on others criteria, for instance, a resistive criterion [1], a power dissipation per unit volume criterion, or other criteria [4]. The complexity of the problem was discussed in details in many reports [4].

However, the common feature of these definitions is that the criteria were chosen to produce "reasonably small" electric power dissipation in practical devices. Based on this the standard approach for describing critical currents for engineering applications is to fit the experimental V-I curves to a power law [1-3], *viz*:

$$V(I) = V_0 + E_c \cdot l \cdot \left(\frac{I}{I_c}\right)^n \tag{1}$$

where, $V_0$ is a fitting parameter related to uncertainties in the experimental data and device sensitivity/noise/electronic drift/etc; $l$ is the conductor length between voltage taps, and $n$ is the power-law exponent. The deduced $I_c$ and $n$-value are key input parameters for the design of engineering applications [1,4-21], and recently a large publicly available data set on $I_c$ of coated conductors of high-$T_c$ YBa$_2$Cu$_3$O$_7$ superconductor (or, 2G-wires) of major wire manufacturers was established [3].

In contrast to this essentially arbitrary convention it was recently shown [22-24] that the *self-field* critical current, $I_c(\text{sf},T)$ is indeed an elementary property, governed only by fundamental superconducting parameters: the London penetration depth, $\lambda$, and (more weakly so) the Ginzburg-Landau parameter, $\kappa = \lambda/\xi$, where $\xi$ is the coherence length. We found that dissipation sets in when the current density, $J_c$, reaches a threshold value of $B_c/\mu_0\lambda$ for type I superconductors or $B_{c1}/\mu_0\lambda$ for type II superconductors. This was established using over 40 data sets for many different thin-film superconductors ranging in thickness from single-atomic-layer [24] to 3 µm [22]. Inferred values of the ground-state penetration depth, $\lambda(0)$, were found to be in excellent agreement with independently-measured values. However, it is important to note that in all cases $I_c$ was determined using the $E_c$. Critical currents defined by the $E_c$ criterion we will designate by $I_{c\_E}$.

It is obvious that these results are in direct opposition to the idea that $E_c$ is merely an arbitrary criterion chosen on engineering convenience. This "customised" $I_c$ definition is perhaps a major obstacle to the idea that the critical current might be a fundamental superconducting property, because parameters such as $\lambda$ inferred from our $J_s$ criterion ultimately depend on an arbitrary electric field criterion. To resolve this problem, we performed self-field critical current measurements on commercial superconducting tapes with the purpose of finding a more fundamental definition for critical current which is not based on the idea of a continuous onset with an arbitrary threshold but, rather, an abrupt and fundamental physical effect accompanying the transition from the dissipation-free to the dissipative regimes.

Experiments were performed on commercial coated conductors and were conducted in the superconducting flux pump facility [10-20] at Robinson Research Institute, Victoria University of Wellington. Self-field magnetic flux density measurements, $B_{surf}(I)$, were performed by immersing the flux pump head and cables (described elsewhere [16-20]) in a liquid nitrogen bath. A 7-element Hall sensor type THV-MOD-1345 (Arepoc s.r.o, Slovakia) was used to record $B_{surf}$ as schematically showed in Fig. 1. Each sensor has an active area of 0.01 mm$^2$ and sensors were spaced 1.5 mm apart. Sensors were placed in position to measure the perpendicular component of the self-field magnetic flux density (Fig. 1).

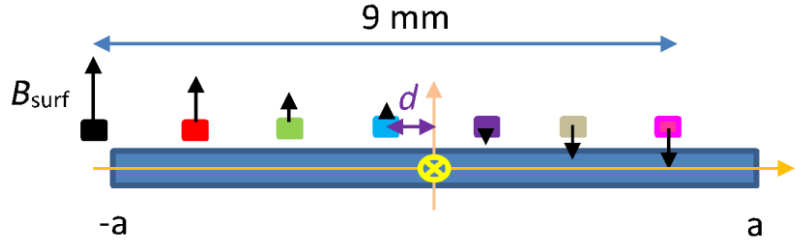

**Figure 1.** Schematic representation of the experiment for 2G wire of width 2*a*. The violet arrow shows the distance, *d*, from the center of the sensor array to the center of the 2G wire.

We designate sensors by the numbers 1, 2, 3, 4, 5, 6, 7 starting from the left (Fig. 1). The manufacturer's measured field and temperature sensitivity calibrations were used for each individual sensor at T = 77.4 K. The width of the sensor array was 9 mm and it was placed at a separation of 0.5 mm from the YBCO side of the 2G wires. We studied wires manufactured by Fujikura Ltd. (width 2*a* = 10 mm, HTS layer thickness 2*b* = 2.3 µm), SuperPower Inc. (2*a* = 12 mm, 2*b* = 1.5 µm), and THEVA (2*a* = 12 mm, 2*b* = 2.6 µm). Transport current was raised in steps of 0.6 A and was held at each step for 3 seconds. The magnetic flux density, $B_{surf}(I)$, was recorded at the end of each current step. Transport current was supplied and recorded using the system which was used for a number of studies [2,11,13,14,17-22], with potential taps placed in the distance of 12 cm.

First set of experiments was performed on Fujikura 2G wire (2*a* = 10 mm, 2*b* = 2.3 µm) for which the critical current defined by the conventional electric-field criterion, $E_c = 1$ µV/cm, (which we designate $I_{c\_E}$) was $I_{c\_E}$ = 464 A and the *n*-value = 35. Fig. 2,a shows the result of $B_{surf}(I)$ measurements for this Fujikura 2G wire for which the center of the sensor array was located at *d* = - 0.3 mm. Thus the entire sensor array was located within the width of the wire and positioned slightly to the left of center (refer to Fig. 1). Sensor 1 was close to the wire edge. Colours for $B_{surf}$ curves in Fig.2,a correspond to color coding of sensors shown in Fig. 1.

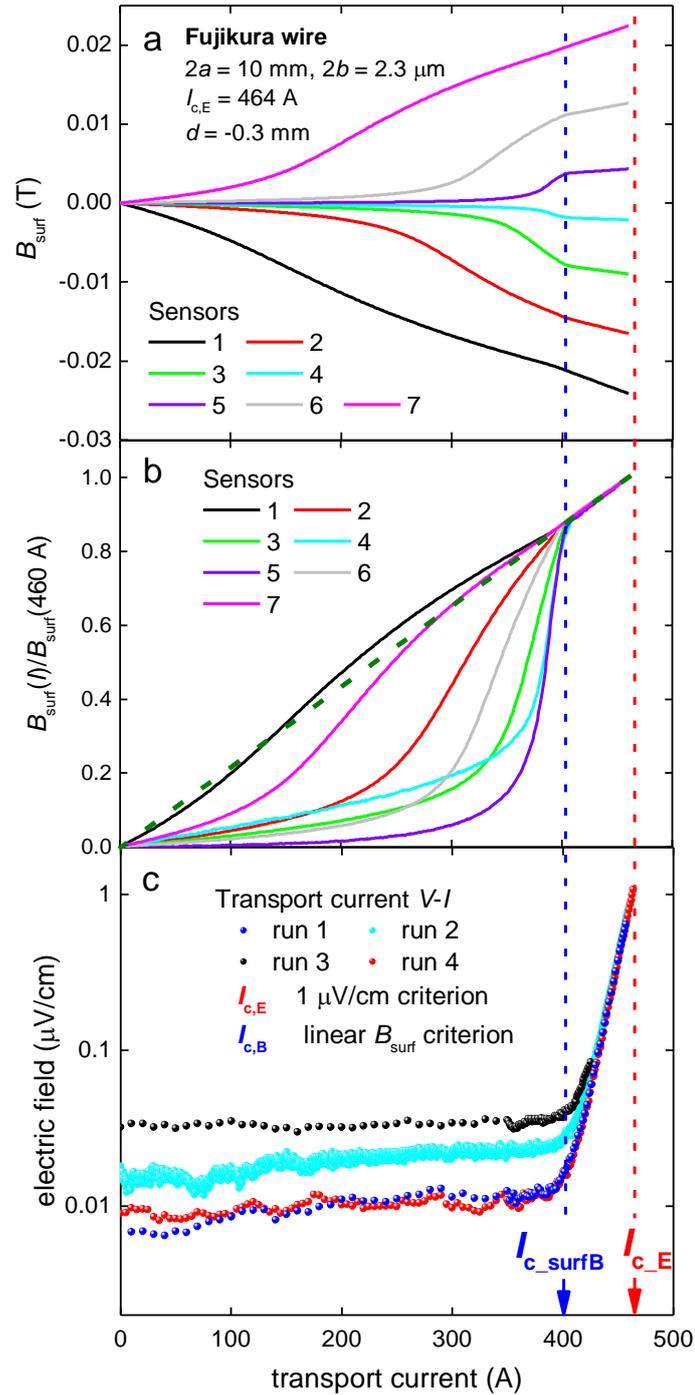

**Figure 2.** Transport current dependence of (a) $B_{surf}(I)$, (b) normalised $B_{surf}(I)/B_{surf}(I = 460A)$ and (c) conventional *V-I* curve for Fujikura 2G wire ($2a = 10$ mm) with sensor array placed at $d = -0.3$ mm. The dashed green line in (b) shows the extrapolation of the linear part of $B_{surf}(I)/B_{surf}(I = 460A)$ back to $I = 0$ A. Note that panel (c) has a logarithmic scale for the vertical axis.

Fig. 2,*b* shows the normalized $B_{surf}(I)/B_{surf}(I = 460$ A) values and Fig. 2,*c* shows the *V-I* curves recorded in four different runs. For each *V-I* curve the background/drift/noise (i.e., $V_0$ in Eq. 1) level differed, but in all cases it did not exceed of 35 nV/cm.

Our most important finding is that as soon as the $V(I)$ curves just start to rise above the background (which is clearly visible only in a logarithmic scale for $V$ due to its very small value) then the surface flux density, $B_{surf}(I)$, for all sensors abruptly changes its dependence from non-linear to linear (Figs. 2,a and 2,b). We will designate this value of the transport current as $I_{c\_surfB}$.

It is important to note that this change occurred at $I_{c\_surfB} = 400$ A, below $I_{c\_E} = 464$ A. This change is most evident in the normalized $B_{surf}(I)/B_{surf}(I = 460$ A$)$ plots in Fig. 2,b, where all curves, which have different shapes for $I < I_{c\_surfB}$, collapse onto a single straight line for $I > I_{c\_surfB}$. Linear extrapolation of this line back to $I = 0$ A intersects the vertical axis at $B = 0$ T with high accuracy (dashed green line in Fig. 2,b). Such linearity is a common characteristic of the resistive state [25], but it is important to note that the conductor is still in the superconducting state.

We need to note, that the transport current distribution for $I < I_{c\_surfB}$ is strongly non-uniform across the wire width and for instance when $I < 0.4*I_{c\_surfB}$ the current dominantly flows near the tape edges (see curves 1,7). Only when $I \sim 0.95 I_{c\_surfB}$ does the central part of the wire start to carry a non-negligible fraction of the total current (see, curves 4,5).

Overall, the data presented in Fig. 2 show that the transition from a dissipation-free to a dissipative regime occurs for the whole 2G-tape simultaneously across the entire width and length as each sensor measures the tape in one particular location only (within an area of 0.01 mm$^2$) while the voltage potentials were placed at a distance of 12 cm. This means that $I_c$ defined as the threshold in $B_{surf}(I)$ none-linearity can be measured at any tape location within the length and width.

The next set of experiments were performed for the SuperPower 2G-wire for which the sensor array was also placed as close as possible to the center of wire ($d$ was estimated to be about 0.1 mm). As the sensor array had a width of 9 mm all sensors were located inside of the wire width. For this SuperPower tape $I_{c\_E}$ = 516 A with $n$-value = 30.5 (Fig. 3).

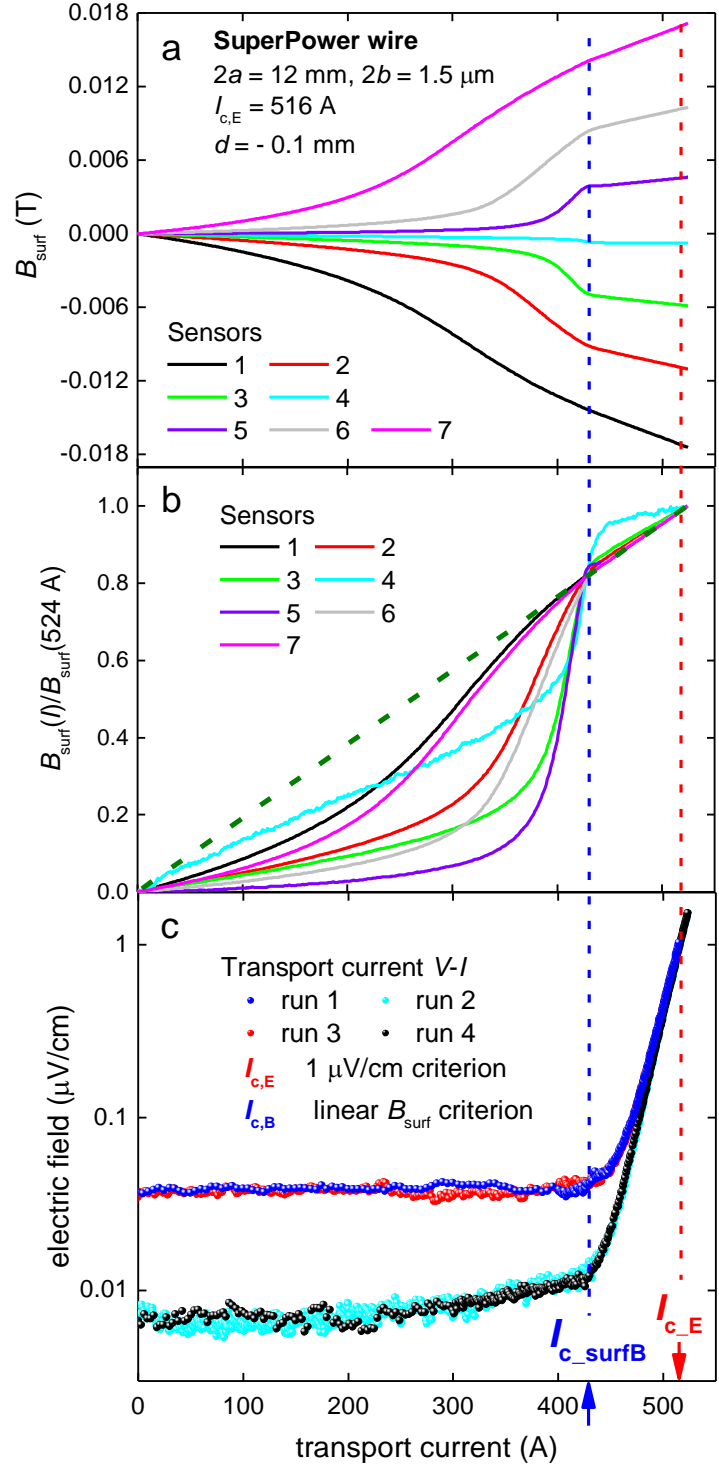

**Figure 3.** Transport current dependence of (a) $B_{surf}(I)$, (b) normalised $B_{surf}(I)/B_{surf}(I = 524$ A$)$ and (c) conventional $V$-$I$ curve (c) for SuperPower 2G wire ($2a$ = 12 mm) with sensor array centered at $d$ = -0.1 mm. The dashed green line in (b) shows the extrapolation of the linear part of $B_{surf}(I)/B_{surf}(I = 524$ A$)$ back to $I$ = 0 A.

Again, as for the Fujikura wire, we found that as soon as the $V(I)$ curves just start to rise above the background (Fig. 3,c) then the surface flux density, $B_{\text{surf}}(I)$, for all sensors changes shape from non-linear to linear behaviour (Figs. 3,a and 3,b). The change in $B_{\text{surf}}(I)$ occurred at $I_{\text{c\_surfB}} = 430$ A, as can be seen both for the absolute $B_{\text{surf}}(I)$ (Fig. 3,a) and the normalized $B_{\text{surf}}(I)/B_{\text{surf}}(I = 524$ A$)$ in Fig. 3,b. All the normalized curves collapsed to the same straight line, similar to Fujikura tape (Fig. 2,b), except for sensor 4. This sensor was positioned so close to the wire center that the absolute value for $B_{\text{surf}}(I)$ was very small and our experimental setup was not able to accurately measure such a small field. On scaling up as in panel (b) any small deviations are excessively amplified.

To demonstrate that the above effects extend to the wires edges and beyond the next experiment was performed for physically the same SuperPower 2G-wire described above. For this the sensor array was placed at the off-centre position of $d = 2.8$ mm so that the last sensor (7) was located outside of the wire while sensor (2) was located 0.2 mm from centre of the wire to avoid the problem experienced in previous experiments of measuring very small field values for sensors close to the wire centre. (Note that due to experimental symmetry $B_{\text{surf}}(I) = 0$ at the wire centre). The experimental results are presented in Fig. 4.

As before, the $V(I)$ curves just commence to rise above the background (Fig. 4,c) as the surface flux density for all sensors changes from non-linear to linear behaviour (Figs. 4,a and 4,b). An interesting result was obtained for Sensor 7 which was situated beyond the wire edge. While the transport current was dissipation-free the shape of $B_{\text{surf}}(I)$ remained slightly non-linear. Simultaneous with the transition for the other sensors (occurring at $I_{\text{c\_surfB}} = 430$ A) we see that $B_{\text{surf}}(I)$ for sensor 7 also starts to be linear. Moreover, $B_{\text{surf}}(I)$ for sensor 7 coincides with $B_{\text{surf}}(I)$ for sensor 5 and both show an excellent match above $I_{\text{c\_surf}B}$. Normalized $B_{\text{surf}}(I)/B_{\text{surf}}(I = 507$ A$)$ lines in Fig. 4,b all collapse to the same straight line with only a small deviation for sensor 2, which was closest to the wire centre and therefore of small absolute magnitude.

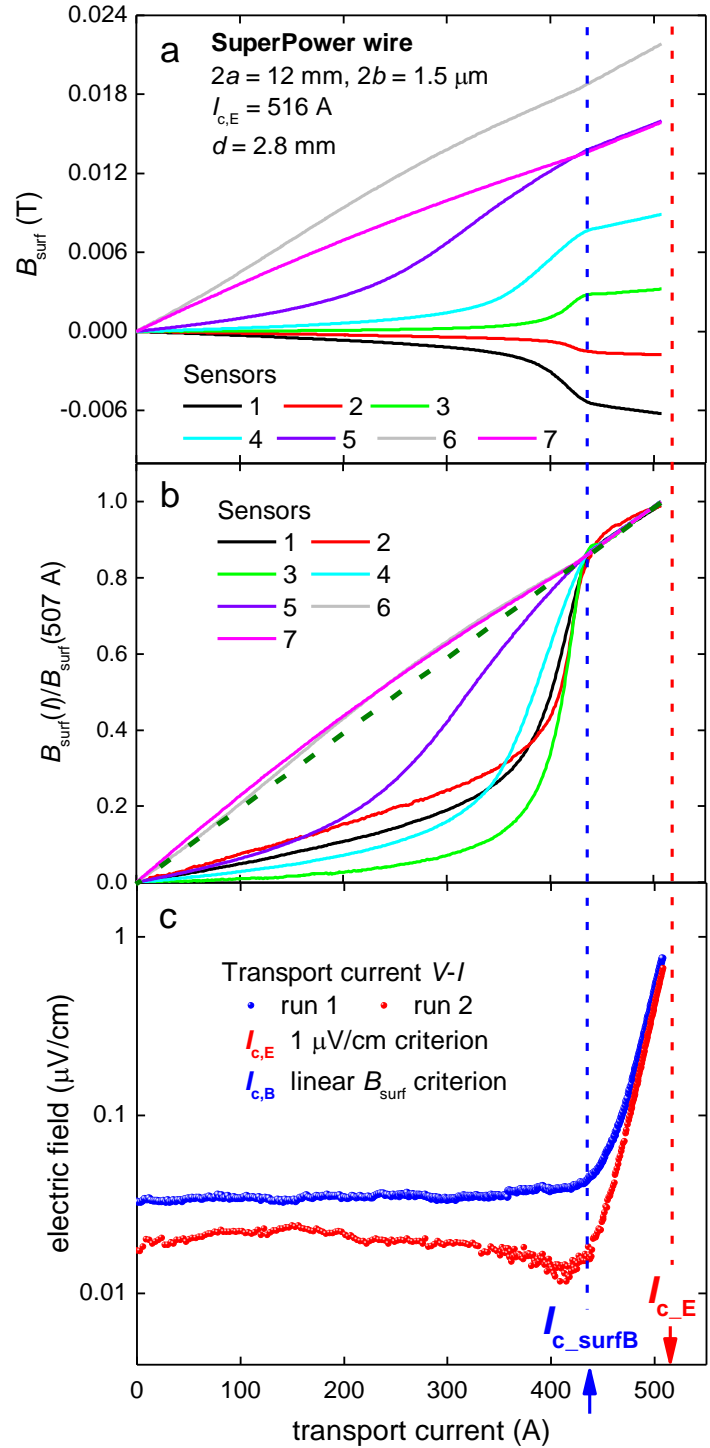

**Figure 4.** Transport current dependence of (a) $B_{surf}(I)$, (b) normalized $B_{surf}(I)/B_{surf}(I = 524$ A$)$ and (c) conventional *V-I* curve for SuperPower 2G-wire ($2a = 12$ mm) with sensor array centred at $d = 2.8$ mm. The dashed green line in (b) shows the extrapolation of the linear part of $B_{surf}(I)/B_{surf}(I = 507$ A$)$ back to $I = 0$ A.

Another set of experiments was performed with the sensor array shifted to $d = 2.4$ mm (and thus, sensor 7 was again positioned outside of the wire) but the transport current was directed in the opposite direction. For this tape $I_{c\_E} = 516$ A with $n$-value = 32. Experimental data is plotted in Fig. 5. The threshold current for the $B_{surf}(I)$ cross-over was found to be $I_{c\_surfB} = 420$ A, again significantly lower than $I_{c\_E}$.

To make further confirmation that described above effect does not relate to certain 2G-wires we used for the study (like, the architecture/deposition techniques/microstructure/etc.) we performed additional experiments on THEVA 2G-wire ($2a = 12$ mm, $2b = 2.6$ μm), for which the sensor array was placed at $d = 2.9$ mm (and thus, sensor 7 was again positioned outside of the wire). For this THEVA tape $I_{c\_E} = 373$ A with $n$-value = 40.5.

Experimental data is plotted in Fig. 6. The threshold current for the $B_{surf}(I)$ cross-over was found to be $I_{c\_surfB} = 317$ A vs $I_{c\_E} = 373$ A.

Thus, our experimental findings are not related to certainly chosen 2G-wires, but it were observe for three different types of 2G-wires manufactured by different deposition techniques and used different substrates/buffer layers/etc.

Here we have measured and analyzed the surface perpendicular magnetic flux density generated by transport currents in second generation YBCO wires with the purpose of finding a fundamental, non-criterion-based definition for critical currents. We observed that there is a distinct threshold value for transport current at which dissipation sets in. At this threshold current, $I_{c\_B}$, two simultaneous process occur:

1. There is an abrupt transition from a non-linear $B_{surf}(I)$ observed at all points on the superconductor surface to a linear dependence.

2. There is an abrupt onset to a non-zero electric field just above the sensitivity of measuring system. This means that $I_{c\_surfB}$ is the fundamental current at which dissipation commences.

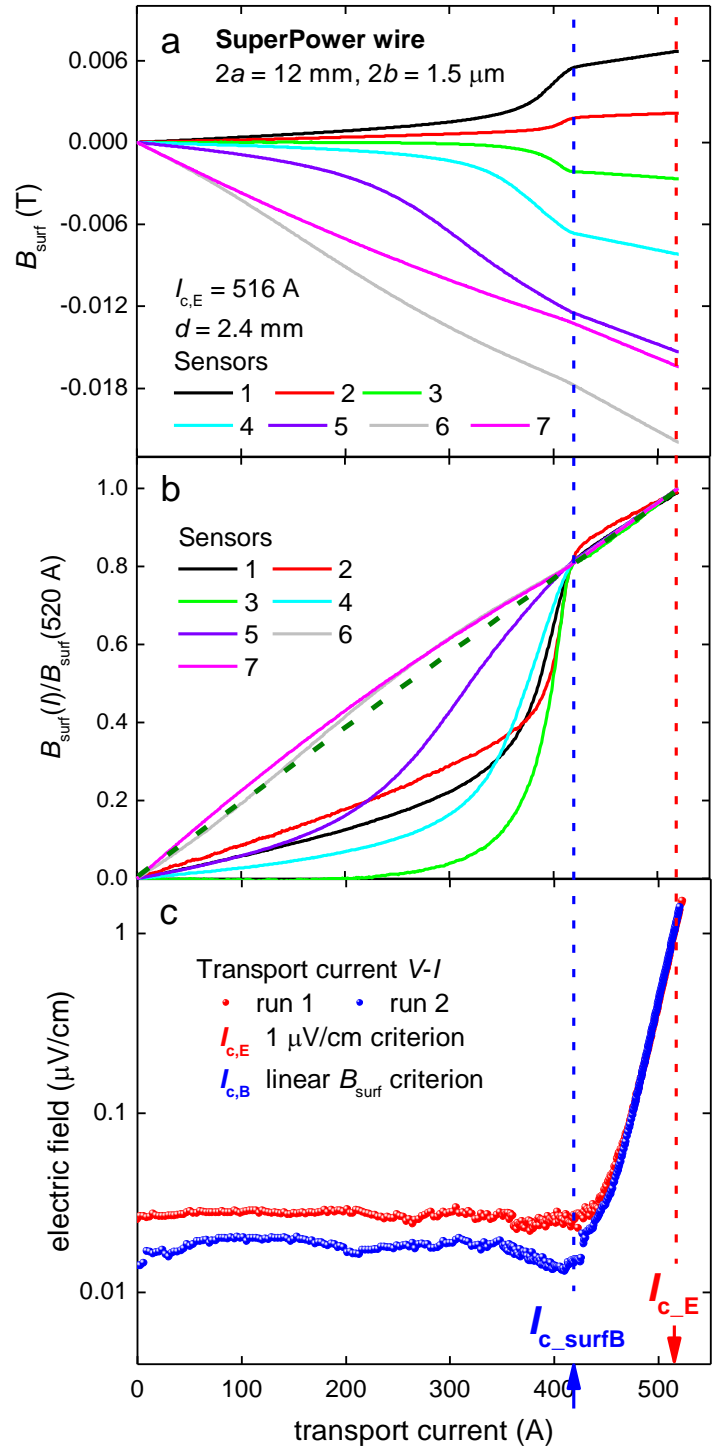

**Figure 5.** Transport current dependence of (a) $B_{\text{surf}}(I)$, (b) normalized $B_{\text{surf}}(I)/B_{\text{surf}}(I = 524 \text{ A})$ and (c) conventional $V$-$I$ curve for SuperPower 2G-wire ($2a = 12$ mm) with sensor array centred at $d = 2.4$ mm. The dashed green line in (b) shows the extrapolation of the linear part of $B_{\text{surf}}(I)/B_{\text{surf}}(I = 520 \text{ A})$ back to $I = 0$ A.

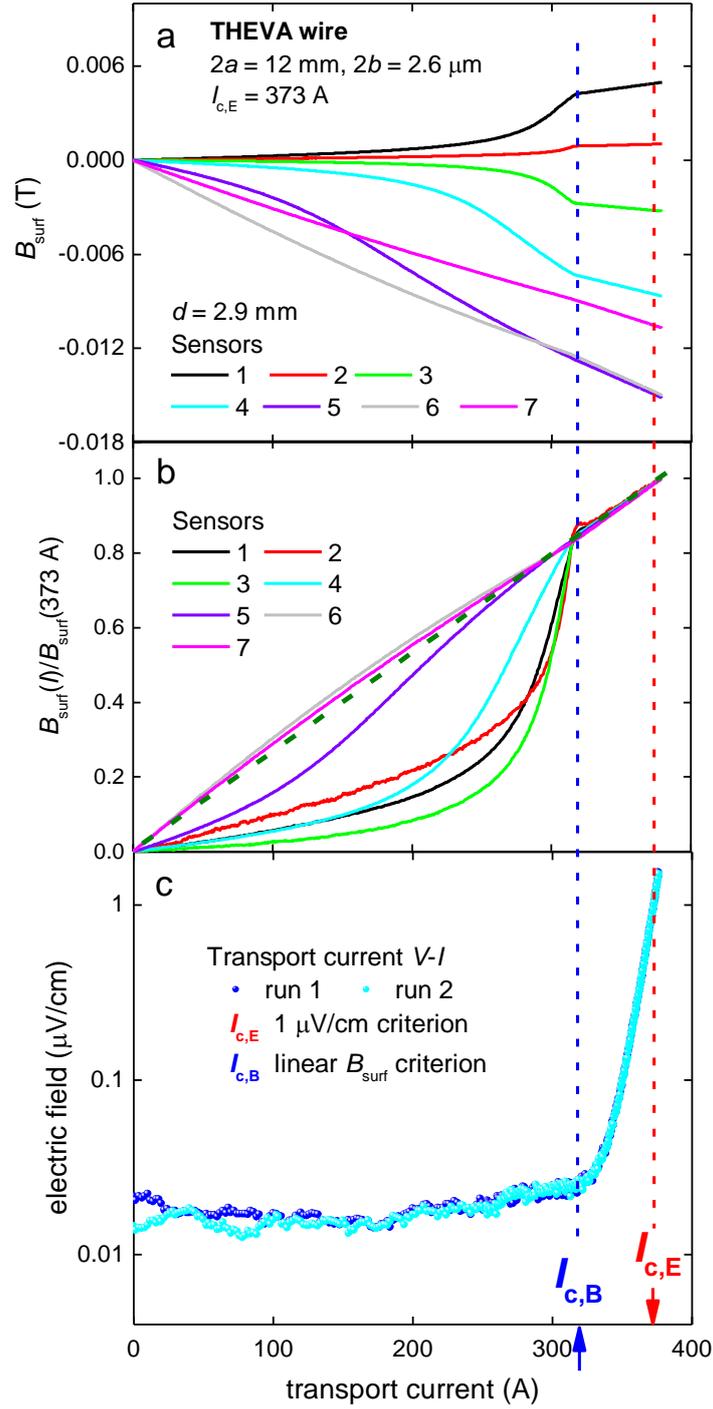

**Figure 6.** Transport current dependence of (a) $B_{surf}(I)$, (b) normalized $B_{surf}(I)/B_{surf}(I = 524\ A)$ and (c) conventional $V$-$I$ curve for THEVA 2G-wire ($2a$ = 12 mm) with sensor array centred at $d$ = 2.9 mm. The dashed green line in (b) shows the extrapolation of the linear part of $B_{surf}(I)/B_{surf}(I = 373\ A)$ back to $I$ = 0 A.

Because each of these occurs simultaneously across the wire width we can conclude that $I_c$ is fundamentally characterised by a transformation for the whole tape, rather than initiation of localised dissipation spots in some weak-linked or high-current density areas. Because of this we propose to designate $I_{c,\text{surfB}}$ as the fundamental critical current, one which is free from arbitrary criterion-based definitions such as electric field, resistance or power dissipation.

We found that $I_{c\_\text{surfB}}$ for YBCO 2G-wires at $T = 77.4$ K is about 8%-15% lower than the conventional critical current measurement, $I_{c\_E}$, determined by the $E_c = 1$ µV/cm criterion. Where previously we have calculated the London penetration depth, $\lambda$, from self-field $J_{c\_E}$ data based on the usual $E_c$ criterion these values would need to be modified. However, as $J_c(\text{sf}) \propto \lambda^{-3}$ [22] for thin samples with $b \leq \lambda$ then deduced $\lambda$ values will be altered (increased) by 3-5% only. Such small changes will not alter any of our previous findings reported in [22-24]. In fact, defining critical current according to $I_{c\_\text{surfB}}$ immediately resolves and removes the obstacle mentioned in the introduction, namely the previous paradox that fundamental parameters of superconductors (such as $\lambda$ and $\Delta$) could be deduced from $J_c$ (sf) values which are measured using an arbitrary, customised electric field criterion. In the above we have shown that there is an underlying more fundamental definition of $J_c(\text{sf})$ which is entirely consistent with the demonstrably fundamental character of this particular property. We need to stress that, the detailed origins of this effect remain be fully established.

Author thanks Dr. A.E. Pantoja and Dr. R.A. Badcock for help and technical support in performing experiments, and Prof. J.L. Tallon for fruitful discussions and help in preparing the manuscript. Author thanks the PBRF Research Support Grant, Victoria University of Wellington No. 215637 and the Marsden Fund of New Zealand Grant Award No. 3648 for financial support.